\documentclass[11pt,superscriptaddress,aps,prd,preprint, showkeys]{revtex4}

\everymath{\displaystyle}
\usepackage{graphicx}

\usepackage{amsmath,amssymb,mathrsfs}

\newcommand{\bea}{\begin{eqnarray}}
\newcommand{\eea}{\end{eqnarray}}

\begin{document}

\title{On Thermodynamics of Kerr Black Hole}

\author{S. C. Ulhoa}\email[]{sc.ulhoa@gmail.com}
\affiliation{International Center of Physics, Instituto de F\'isica, Universidade de Bras\'ilia, 70910-900, Bras\'ilia, DF, Brazil} \affiliation{Canadian Quantum Research Center,\\ 
204-3002 32 Ave Vernon, BC V1T 2L7  Canada} 

\author{A. F. Santos}\email[]{alesandroferreira@fisica.ufmt.br}
\affiliation{Instituto de F\'{\i}sica, Universidade Federal de Mato Grosso,\\
78060-900, Cuiab\'{a}, Mato Grosso, Brazil}

\author{E. P. Spaniol}
\email{spaniol.ep@gmail.com} \affiliation{UDF Centro Universit\'ario
and Centro Universit\'ario de Bras\'ilia UniCEUB, Bras\'ilia, DF,
Brazil.}

\author{Faqir C. Khanna\footnote{Professor Emeritus - Physics Department, Theoretical Physics Institute, University of Alberta\\
Edmonton, Alberta, Canada}}\email[]{khannaf@uvic.ca}
\affiliation{Department of Physics and Astronomy, University of Victoria,\\
3800 Finnerty Road Victoria, BC, Canada}

\begin{abstract}

The gravitational Stefan-Boltzmann law is considered for the Kerr black hole in the weak-field limit.  The energy-momentum tensor predicted by Teleparallelism Equivalent to General Relativity (TEGR) is used in the Thermo Field Dynamics (TFD) formalism to thermalize the field. A temperature-dependent gravitational pressure is obtained. Regions of divergent heat capacity are observed. According to Landau theory, it allows the existence of distinct phases around the Kerr black hole. 

\end{abstract}

\maketitle

\date{\today}
\section{Introduction} \label{sec.1}

General relativity is fascinating both for its geometric formulation, which changed the physical view of fundamental interactions, and for its predictions.  Among the latter, there are black holes, which are the densest objects in the Universe and also the most mysterious \cite{dinverno}.  It was quickly realized that the first solutions of Einstein's equations could describe these objects.  Einstein himself doubted the existence of black holes, despite being legitimate solutions of his equations.  The main argument raised was the problem of the non-removable singularity of black holes and, although there are strong indications of their existence, there are still those who doubt it.  Recently, photographs supposedly of several black holes support their existence, as well as the measurements of gravitational waves arising from the coalescence of two black holes \cite{waves,photo}.  It is interesting to note that the no-hair theorem states that black holes must be described by fundamental quantities such as mass, charge, and angular momentum. It is therefore expected that even a thermodynamic description of black holes will respect such a theorem, such as Hawking's thermodynamics.  In this approach, the thermodynamic state is limited to a region of space, such as the area of the event horizon of the black hole. Furthermore, there is no guarantee that the temperature, constructed from a geometric structure, will interact with a thermometer and allow one to confirm the predictions of the theory \cite{hawking, Bekenstein, Jacobson, Padmanabhan, Isi, Wald}. The  Hawking approach applied to Kerr black hole does not take into account the pressure term in the first law of thermodynamics, at least not explicitly \cite{dolan}.  Therefore, we believe that this approach is not complete.  It is necessary to treat temperature as an independent variable in black hole thermodynamics, as it may be linked to the internal structure of the black hole, rather than its geometry.

One of the first attempts to introduce temperature in field theory is due to the Matsubara approach \cite{matsubara}. This formalism associates time with temperature, thus it leads to the loss of the temporal evolution of the field. Another approach is to introduce the Thermo Field Dynamics (TFD)~\cite{khanna}.  In this approach, the temperature coexists with the temporal coordinate. Due to the TFD topological structure, different phenomena, such as the Stefan-Boltzmann law and the Casimir effect, are analyzed on an equal footing.  In order to use TFD, it is necessary to known two quantities, namely, the energy-momentum tensor and the propagator of the field.  Then, for gravitation it is necessary to search for an alternative formulation of the field, given the well-known problem of defining gravitational energy in general relativity.  Therefore, the formulation of the gravitational field will be considered in terms of Teleparallelism Equivalent to General Relativity (TEGR)~\cite{maluf}.  In this approach the energy-momentum tensor is well defined and allows the use of TFD in terms of weak field approximation~\cite{flatTFD}.  These expressions are applied to the Kerr black hole to deal with the gravitational thermodynamics in this regime.  Then,  the energy of the Stefan-Boltzmann law is obtained, which allows the calculation of gravitational pressure via the first law of thermodynamics, which leads to the entropy and thermal capacity.

This paper is organized as follows. In section II, ideas about TFD are considered. In section III, the TEGR is introduced to obtain two quantities, the Green function for gravity and energy-momentum tensor. In section IV, the expressions for Kerr black hole metric are applied to calculate Stefan-Boltzmann law, pressure, entropy and gravitational heat capacity. Conclusions are presented in the last section. Natural units are used,  i.e. $G=c=k_B=1$.

\section{Thermo Field Dynamics (TFD)} \label{sec.2}

TFD is a real-time approach to introducing the effects of temperature into a quantum field theory  \cite{Umezawa1, Umezawa2, Khanna0, Umezawa22, Khanna1, Khanna2, GBT}. In this formalism, a thermal ground state is constructed. For such a construction, two elements are needed: the duplication of the Fock space, i.e. ${\cal S}_T={\cal S}\otimes \tilde{\cal S}$, where ${\cal S}$ is the original Fock space and $\tilde{\cal S}$ is the dual Fock space, and the Bogoliubov transformation. An operator in the original Fock space is associated with two operators in the thermal Fock space, one in tilde space and the other in non-tilde space. The tilde and non-tilde operators are related through the tilde conjugation rules. These rules are defined as
\bea
({\cal A}_i{\cal A}_j)^\thicksim &=& \tilde{{\cal A}_i}\tilde{{\cal A}_j},\nonumber\\(c{\cal A}_i+{\cal A}_j)^\thicksim &=& c^*\tilde{{\cal A}_i}+\tilde{{\cal A}_j}, \nonumber\\ ({\cal A}_i^\dagger)^\thicksim &=& \tilde{{\cal A}_i}^\dagger, \nonumber\\ (\tilde{{\cal A}_i})^\thicksim &=& -\xi {\cal A}_i,
\eea
where ${\cal A}$ is an arbitrary operator, $\xi = -1$ for bosons and $\xi = +1$ for fermions. 

The Bogoliubov transformation introduces thermal effects by applying a rotation between the tilde and non-tilde variables. To illustrate how this transformation works, let's consider $\tilde{d}^\dagger$ as creation operator and $d$ as destruction operator. Then
\bea
\left( \begin{array}{cc} d(\alpha, k)  \\ \tilde d^\dagger(\alpha, k) \end{array} \right)={\cal U}(\alpha)\left( \begin{array}{cc} d(k)  \\ \tilde d^\dagger(k) \end{array} \right),
\eea
where ${\cal U}(\alpha)$ is the Bogoliubov transformation that is defined as
\bea
{\cal U}(\alpha)=\left( \begin{array}{cc} u(\alpha) & -v(\alpha) \\
\xi v(\alpha) & u(\alpha) \end{array} \right).
\eea
The $\alpha$ parameter is defined as $\alpha=(\alpha_0,\alpha_1,\cdots\alpha_{D-1})$. The temperature effect is introduced choosing  $\alpha_0\equiv\beta$ and $\alpha_1,\cdots\alpha_{D-1}=0$, with $\beta=1/T$. Since TFD has a topological structure, the $\alpha$ parameter is called the compactification parameter. Any manifold dimension can be compactified into a hyper-circumference of length $\alpha$. Therefore, temporal or spatial compactifications, or both together, make it possible to investigate different phenomena on an equal footing.

An interesting point is to analyze the Green function or propagator of a quantum field in the TFD approach. Since it will be used in the next section, let us consider the propagator of the scalar field given as
\bea
G_0^{(ab)}(x-x';\alpha)=i\langle 0,\tilde{0}| \tau[\phi^a(x;\alpha)\phi^b(x';\alpha)]| 0,\tilde{0}\rangle,
\eea
where $a,b$ $=1,2$, and $\tau$ is the time ordering operator. The $\alpha$ parameter is introduced through the transformation $\phi(x;\alpha)={\cal U}(\alpha)\phi(x){\cal U}^{-1}(\alpha)$. The Fourier transform of the propagator is
\bea
G_0^{(ab)}(x-x';\alpha)=i\int \frac{d^4k}{(2\pi)^4}e^{-ik(x-x')}G_0^{(ab)}(k;\alpha)
\eea
with $G_0^{(ab)}(k;\alpha)={\cal U}(\alpha)G_0^{(ab)}(k){\cal U}^{-1}(\alpha).$ Here, only the non-tilde part of the Green function is considered, i.e.
\bea
G_0^{(11)}(k;\alpha)=G_0(k)+ v^2(k;\alpha)[G^*_0(k)-G_0(k)].
\eea
It is important to note that, there is a similar function for the tilde part. The function $v^2(k;\alpha)$ is the generalized Bogoliubov transformation defined as
\bea
v^2(k;\alpha)&=&\sum_{s=1}^p\sum_{\lbrace\sigma_s\rbrace}2^{s-1}\sum_{l_{\sigma_1},...,l_{\sigma_s}=1}^\infty(-\xi)^{s+\sum_{r=1}^sl_{\sigma_r}}\exp\left[{-\sum_{j=1}^s\alpha_{\sigma_j} l_{\sigma_j} k^{\sigma_j}}\right],
\eea
where $p$ is the number of compactified dimensions, $\lbrace\sigma_s\rbrace$ is the set of all combinations with $s$ elements and $k$ is the 4-momentum.

\section{Teleparallel Gravity} \label{sec.3}

In this section, the Teleparallelism Equivalent to General Relativity (TEGR) \cite{maluf} is briefly presented. The equivalence between General Relativity (GR) and TEGR takes place at the dynamic level, despite the dynamic variables of both theories being different.  In the first case it is the metric tensor, while in the second, the tetrad field.  These variables are related by $e^{a}\,_{\mu} e_{a\nu} =g_{\mu\nu}$. The tetrad field relates two symmetries: on the one hand the Latin indices represent the Lorentz transformations, $a=(0)...(3)$, on the other the Greek indices are used for diffeomorphisms, $\mu=0..3$.  Furthermore, the tetrad field is adapted to an observer.  The arbitrariness of the choice of frame of reference is intrinsic to the tetrad field as it has 16 independent components.  That is, 6 components more than the metric tensor.  So there are a myriad of possible tetrads for each metric tensor. The TEGR is built in a Weitzenb\"ock space, in such geometry the connection has zero curvature and dynamic torsion.  It is interesting to note that this picture is the opposite of what exists in Riemannian geometry.  Despite this, there is a relationship between the two, which is precisely the core of the dynamic equivalence between GR and TEGR. The Weitzenb\"ock connection is
$$\Gamma_{\mu\lambda\nu}=e^{a}\,_{\mu}\partial_{\lambda}e_{a\nu}\,,$$
which has the following non-vanishing torsion
\begin{equation}
T^{a}\,_{\lambda\nu}=\partial_{\lambda} e^{a}\,_{\nu}-\partial_{\nu}
e^{a}\,_{\lambda}\,. \label{4}
\end{equation}

Christoffel's symbols, which is the connection of a Riemannian geometry, are related to the Weitzenb\"ock connection, through the following identity
\begin{equation}
\Gamma_{\mu \lambda\nu}= {}^0\Gamma_{\mu \lambda\nu}+ K_{\mu
\lambda\nu}\,, \label{2}
\end{equation}
where 
\begin{eqnarray}
K_{\mu\lambda\nu}&=&\frac{1}{2}(T_{\lambda\mu\nu}+T_{\nu\lambda\mu}+T_{\mu\lambda\nu})\,,\label{3}
\end{eqnarray}
is the contortion tensor. Using this mathematical identity, the following relationship is obtained
\begin{equation}
eR(e)\equiv -e(\frac{1}{4}T^{abc}T_{abc}+\frac{1}{2}T^{abc}T_{bac}-T^aT_a)+2\partial_\mu(eT^\mu)\,,\label{eq5}
\end{equation}
where the left side of the equation is the known Hilbert-Einstein Lagrangian density used in GR.  Thus the curvature scalar of Riemannian geometry is related to a quadratic combination of the torsion tensor defined in Weitzenb\"ock space. Then, disregarding the total divergence, the TEGR Lagrangian density is written as
\begin{eqnarray}
\mathfrak{L}(e_{a\mu})&=& -\kappa\,e\,(\frac{1}{4}T^{abc}T_{abc}+
\frac{1}{2} T^{abc}T_{bac} -T^aT_a) -\mathfrak{L}_M\nonumber \\
&\equiv&-\kappa\,e \Sigma^{abc}T_{abc} -\mathfrak{L}_M\; ,\label{6}
\end{eqnarray}
where $\mathfrak{L}_M $ stands for the matter fields Lagrangian density and the tensor $\Sigma^{abc}$ is defined by 
\begin{equation}
\Sigma^{abc}=\frac{1}{4} (T^{abc}+T^{bac}-T^{cab}) +\frac{1}{2}(
\eta^{ac}T^b-\eta^{ab}T^c)\;. \label{7}
\end{equation}
It is interesting to note that Lagrangian, Eq. (\ref{6}), is invariant by coordinate transformations and Lorentz global transformations.  It is not invariant by local Lorentz transformations due to the exclusion of the total divergence in Eq. (\ref{eq5}), which is invariant by Lorentz (global and local) and coordinate transformations.

By varying the TEGR Lagrangian density with respect to the tetrad field, the following equation is obtained
\begin{equation}
\partial_\nu\left(e\Sigma^{a\lambda\nu}\right)=\frac{1}{4\kappa}
e\, e^a\,_\mu( t^{\lambda \mu} + T^{\lambda \mu})\;, \label{10}
\end{equation}
where
\begin{equation}
t^{\lambda \mu}=\kappa\left[4\,\Sigma^{bc\lambda}T_{bc}\,^\mu- g^{\lambda
\mu}\, \Sigma^{abc}T_{abc}\right]\,, \label{11}
\end{equation}
is the gravitational energy-momentum tensor, while $T^{\lambda \mu}$ refers to the matter fields. 

It should be noted that the $\Sigma^{a\lambda\nu}$ tensor is anti-symmetric in the last two indices, this implies the following identity

\begin{equation}
\partial_\lambda\partial_\nu\left(e\Sigma^{a\lambda\nu}\right)\equiv0\,.\label{12}
\end{equation}
With that, a conservation law is built for such a quantity. This leads to the following definition for the energy-momentum vector
\begin{equation}
P^a = \int_V d^3x \,e\,e^a\,_\mu(t^{0\mu}+ T^{0\mu})\,, \label{14}
\end{equation}
or
\begin{equation}
P^a =4k\, \int_V d^3x \,\partial_\nu\left(e\,\Sigma^{a0\nu}\right)\,. \label{14.1}
\end{equation}
Then, an energy-momentum vector that is invariant by coordinate transformations is constructed. However, it depends on the choice of the frame of reference, as in special relativity. Such characteristics make the quantity defined within the TEGR desirable and resilient to theoretical or experimental consistency tests.

Once the energy-momentum vector is well defined, to achieve the objective of implementing the TFD in the TEGR framework it is necessary to write the field propagator.  This is possible assuming the weak field approximation, which in terms of the metric tensor means
\begin{equation}
g_{\mu\nu}=\eta_{\mu\nu}+h_{\mu\nu},
\end{equation}
which together with expression (\ref{6}), yields the graviton propagator \cite{flatTFD}
\begin{equation}
\langle e_{b\lambda}, e_{d\gamma} \rangle=\Delta_{bd\lambda\gamma} = \frac{\eta_{bd}}{\kappa q^{\lambda} q^{\gamma}}.
\end{equation}
Hence the Green function is
\bea
G_0(x,x')=-i\Delta_{bd\lambda\gamma}\,g^{\lambda\gamma}\eta^{bd}.
\eea
Explicitly it is given by
\begin{equation}
G_0(x,x')= -\frac{i64\pi}{q^{2}}\,,
\end{equation}
with $q=x-x'$, where $x$ and $x'$ are four vectors. In such an approximation the gravitational energy-momentum tensor $t^{\lambda \mu}$ becomes
\begin{eqnarray}
t^{\lambda\mu}(x) &=& \kappa\Bigl[g^{\mu\alpha}\partial^{\gamma}e^{b\lambda}\partial_{\gamma}e_{b\alpha} - g^{\mu\gamma}\partial^{\alpha}e^{b\lambda}\partial_{\gamma}e_{b\alpha} - g^{\mu\alpha}(\partial^{\lambda}e^{b\gamma}\partial_{\gamma}e_{b\alpha} - \partial^{\lambda}e^{b\gamma}\partial_{\alpha}e_{b\gamma})\nonumber\\
        & &-2g^{\lambda\mu}\partial^{\gamma}e^{b\alpha}(\partial_{\gamma}e_{b\alpha}-\partial_{\alpha}e_{b\gamma})\Bigl]\,.
\end{eqnarray}
Therefore, the vacuum expectation value of the gravitational energy-momentum tensor is 
\bea
\langle t^{\lambda\mu}(x)\rangle &=& \langle 0|t^{\lambda\mu}(x)|0\rangle,\nonumber\\
&=& \lim_{x'^\mu\rightarrow x^\mu} 4i\kappa\left(-5g^{\lambda\mu}\partial'^{\gamma}\partial_{\gamma} +2g^{\mu\alpha}\partial'^{\lambda}\partial_{\alpha}\right)G_{0}(x-x')\,,\label{em}
\eea
where $\langle e_{c}^{\,\,\,\lambda}(x), e_{b\alpha}(x') \rangle = i\eta_{cb}\,\delta^{\lambda}_{\alpha}\,G_{0}(x-x')$. It is interesting to note that TFD applied to gravitation is a quantum theory for this field.  Then, the mean value calculated above is an observable of the theory obtained from a quantum state at zero temperature.  The temperature is introduced from Bogoliubov transformations.  This means that the quantum state is thermalized and the energy density or momenta is obtained at finite temperature. It is worth noting that the average of the energy-momentum tensor is calculated in terms of the components of the metric tensor. As this result was obtained in the weak-field approximation, this implies that the tetrad satisfies the condition $e_{(0)}\,^i=0$, that is, such a tetrad field is adapted to a stationary reference frame at spatial infinity. It is worth noting that the TEGR is crucial for the thermal description of the gravitational field, since the concept of gravitational energy is well defined in the framework of such a theory. Thus the thermalization of the GR using TFD is not feasible, which highlights the need to use an approach in which the energy-momentum tensor has a physical prediction. Thus, in TEGR, there is a temperature associated with the gravitational field itself, i.e., a gravitational thermodynamics is linked to the measurement of the heat of the field. There are certainly several ways to construct such thermodynamics, but the temperature in all cases must be a parameter that measures the thermal energy of the field, for instance, in reference \cite{ulhoa}, the thermodynamics of a PP gravitational wave is studied by a classic approach.

\section{Thermodynamics of Kerr Black Hole} \label{sec.4}

The Kerr solution in terms of the Boyer-Lindquist coordinates is given by
\begin{eqnarray}
ds^2&=&
-{{\psi^2}\over {\rho^2}}dt^2-{{2\chi\sin^2\theta}\over{\rho^2}}
\,d\phi\,dt
+{{\rho^2}\over {\Delta}}dr^2 +\rho^2d\theta^2+ {{\Sigma^2\sin^2\theta}\over{\rho^2}}d\phi^2\,,
\label{22}
\end{eqnarray}
with the following definitions
\begin{eqnarray}
\Delta&=& r^2+a^2-2Mr\,,  \nonumber \\
\rho^2&=& r^2+a^2\cos^2\theta \,,  \nonumber \\
\Sigma^2&=&(r^2+a^2)^2-\Delta a^2\sin^2\theta\,,  \nonumber \\
\psi^2&=&\Delta - a^2 \sin^2\theta\,, \nonumber \\
\chi &=&2aMr\,.
\label{23}
\end{eqnarray}

Taking the limits $\frac{M}{r}\ll1$ and $\frac{a}{r}\ll1$, the Kerr solution becomes
\bea
ds^{2}=-\left(1-\frac{2M}{r}\right)dt^{2}+\left(1+\frac{2M}{r}\right)dr^{2}+r^{2}d\Omega^{2}-\frac{4J}{r}\sin^{2}\theta dtd\phi,\label{Kerr}
\eea
where $J= Ma$.

In order to investigate some applications at non-zero temperature using Eq. (\ref{Kerr}), let us write the energy-momentum tensor Eq. (\ref{em}) in the TFD notation, i.e.
\bea
\langle t^{\lambda\mu(ab)}(x;\alpha)\rangle = \lim_{x'^\mu\rightarrow x^\mu} 4i\kappa\left(-5g^{\lambda\mu}\partial'^{\gamma}\partial_{\gamma} +2g^{\mu\alpha}\partial'^{\lambda}\partial_{\alpha}\right)G_{0}^{(ab)}(x-x';\alpha).
\eea

Since the energy-momentum is divergent at zero or non-zero temperature, a renormalization prescription is needed. A finite quantity is obtained by making 
\bea
{\cal T}^{\lambda\mu (ab)}(x;\alpha)\equiv \langle t^{\lambda\mu(ab)}(x;\alpha)\rangle-\langle t^{\lambda\mu}(x)\rangle, 
\eea
which leads to
\bea
{\cal T}^{\lambda\mu (ab)}(x;\alpha)=\lim_{x'^\mu\rightarrow x^\mu} 4i\kappa\left(-5g^{\lambda\mu}\partial'^{\gamma}\partial_{\gamma} +2g^{\mu\alpha}\partial'^{\lambda}\partial_{\alpha}\right)\overline{G}_0^{(ab)}(x-x';\alpha)\label{eq30}
\eea
with
\bea
\overline{G}_0^{(ab)}(x-x';\alpha)&=&G_0^{(ab)}(x-x';\alpha)-G_0^{(ab)}(x-x').
\eea

Using Eq. (\ref{Kerr}) the components of Eq. (\ref{eq30}) for $\mu=\nu=0,1,2,3$ are given as
\begin{eqnarray}
{\cal T}^{00 (11)}(x;\alpha) &=&\lim_{x'^\mu\rightarrow x^\mu}\frac{4i \kappa}{\left[\left(1-\frac{2M}{r}\right)+\frac{4J^{2}}{r^{4}}\sin\theta\right]^{2}}
\Biggl\{-3 \partial_{0}^{\prime}\partial_{0}-\frac{12J}{r^{3}}\partial_{0}^{\prime}\partial_{3}+5\Biggl[\frac{1}{\left(1+\frac{2M}{r}\right)}\partial_{1}^{\prime}\partial_{1}\nonumber\\
&+&\frac{1}{r^{2}}\partial_{2}^{\prime}\partial_{2}\Biggl]+\left[\frac{8J^2}{r^{6}}+\frac{5\left(1-\frac{2M}{r}\right)}{r^{2}\sin^{2}\theta}\right]\partial_{3}^{\prime}\partial_{3} \Biggr\}\overline{G}_0^{(11)}(x-x';\alpha),
\end{eqnarray}

\begin{eqnarray}
{\cal T}^{11 (11)}(x;\alpha) &=&	\lim_{x'^\mu\rightarrow x^\mu}\frac{4i \kappa}{\left(1+\frac{2M}{r}\right)}\Biggl\{\frac{-5}{\left[\left(1-\frac{2M}{r}\right)+\frac{4J^{2}}{r^{4}}\sin\theta\right]} \left[-\partial_{0}^{\prime}\partial_{0}-\frac{4J}{r^{3}}\partial_{0}^{\prime}\partial_{3}+\frac{\left(1-\frac{2M}{r}\right)}{r^{2}\sin^{2}\theta}\partial_{3}^{\prime}\partial_{3}\right]\nonumber\\
&-&\frac{5}{r^{2}}\partial_{2}^{\prime}\partial_{2}-\frac{3}{\left(1+\frac{2M}{r}\right)}\partial_{1}^{\prime}\partial_{1}\Biggr\}\overline{G}_0^{(11)}(x-x';\alpha),
\end{eqnarray}

\begin{eqnarray}
{\cal T}^{22 (11)}(x;\alpha) &=&\lim_{x'^\mu\rightarrow x^\mu}\frac{4i \kappa}{r^2}\Biggl\{\frac{5}{\left[\left(1-\frac{2M}{r}\right)+\frac{4J^{2}}{r^{4}}\sin\theta\right]} \left[\partial_{0}^{\prime}\partial_{0}+\frac{4J}{r^{3}}\partial_{0}^{\prime}\partial_{3}-\frac{\left(1-\frac{2M}{r}\right)}{r^{2}\sin^{2}\theta}\partial_{3}^{\prime}\partial_{3}\right]\nonumber\\
&-&\frac{3}{r^{2}}\partial_{2}^{\prime}\partial_{2}-\frac{5}{\left(1+\frac{2M}{r}\right)}\partial_{1}^{\prime}\partial_{1}\Biggr\}\overline{G}_0^{(11)}(x-x';\alpha),
\end{eqnarray}

\begin{eqnarray}
{\cal T}^{33 (11)}(x;\alpha) &=&	\lim_{x'^\mu\rightarrow x^\mu}\frac{4i \kappa}{r^{2}\sin^{2}\theta\left[\left(1-\frac{2M}{r}\right)+\frac{4J^{2}}{r^{4}}\sin\theta\right]^{2}}\Biggl\{\left[5 \left(1-\frac{2M}{r}\right)+ \frac{8J^2 \sin^{2}\theta}{r^{4}}\right]\partial_{0}^{\prime}\partial_{0}\nonumber\\
&-&5\left(1-\frac{2M}{r}\right)\left[\frac{1}{\left(1+\frac{2M}{r}\right)} \partial_{1}^{\prime}\partial_{1}+\frac{1}{r^2} \partial_{2}^{\prime}\partial_{2}\right]+ \frac{12J \left(1-\frac{2M}{r}\right)}{r^3}\partial_{0}^{\prime}\partial_{3}\nonumber\\
&-&\frac{3 \left(1-\frac{2M}{r}\right)^2}{r^2}\partial_{3}^{\prime}\partial_{3}\Biggr\}\overline{G}_0^{(11)}(x-x';\alpha).
\end{eqnarray}

Now, using these results, the Stefan-Boltzmann law associated with the gravitational field is calculated and some thermodynamics quantities are analyzed.

\subsection{Gravitational Stefan-Boltzmann law for Kerr Black Hole}

To obtain the effect of temperature related to the gravitational field, a particular topology is considered. For this proposal, $\alpha=(\beta,0,0,0)$ is chosen. This means that in this topology the temporal coordinate is compactified into a circumference of length $\beta$.
For this case, the generalized Bogoliubov transformation takes the form
\bea
v^2(\beta)=\sum_{l_0=1}^{\infty}e^{-\beta k^0l_0}\label{BT1}
\eea
and the Green function is given as
\bea
\overline{G}_0(x-x';\beta)=2\sum_{l_0=1}^{\infty}G_0(x-x'-i\beta l_0n_0),\label{GF1}
\eea
with $n_0=(1,0,0,0)$, being a time-like vector. Then the component with $\mu=\nu=0$ of the energy-momentum tensor becomes
\begin{eqnarray}
{\cal T}^{00 (11)}(x;\beta) &=&2\lim_{x'^\mu\rightarrow x^\mu}\sum_{l_0=1}^{\infty}\frac{4i \kappa}{\left[\left(1-\frac{2M}{r}\right)+\frac{4J^{2}}{r^{4}}\sin\theta\right]^{2}}
\Biggl\{-3 \partial_{0}^{\prime}\partial_{0}-\frac{12J}{r^{3}}\partial_{0}^{\prime}\partial_{3}+5\Biggl[\frac{1}{\left(1+\frac{2M}{r}\right)}\partial_{1}^{\prime}\partial_{1}\nonumber\\
&+&\frac{1}{r^{2}}\partial_{2}^{\prime}\partial_{2}\Biggl]+\left[\frac{8J^2}{r^{6}}+\frac{5\left(1-\frac{2M}{r}\right)}{r^{2}\sin^{2}\theta}\right]\partial_{3}^{\prime}\partial_{3} \Biggr\}G_0(x-x'-i\beta l_0n_0).
\end{eqnarray}
Performing the sum and after some calculations, this component takes the form
\bea
{\cal T}^{00 (11)}(T) &=&\frac{32\pi^4\, T^4}{45(2M-r)^3(2Mr^3-r^4-4J^2\sin^2\theta)^2}\Bigl[8M^2r^9+2Mr^{10}-3r^{11}+88J^2Mr^6\sin^2\theta\nonumber\\
&-&44J^2r^7\sin^2\theta-72J^2Mr^9\sin^2\theta+36J^2r^{10}\sin^2\theta-64J^4r^3\sin^4\theta\Bigl].\label{E}
\eea
This is the gravitational Stefan-Boltzmann law for the Kerr black hole.

\subsection{The First Law of Thermodynamics}

It should be noted that the expression (\ref{E}) is precisely the energy density, i.e. ${\cal T}^{00 (11)}= \frac{\partial U}{\partial V} $, where U is the energy and V is the volume.  On the other hand, there is no clear identification of how to establish the pressure, as it is a combination of the respective spatial components of the energy-momentum tensor.  Then an expression for the pressure is derived from the first law of thermodynamics, i.e.
\begin{equation}\label{primeiralei}
\left(\frac{\partial U}{\partial V}\right)_{T}=T\left(\frac{\partial P}{\partial T}\right)_{V}-P\,.
\end{equation}
Solving this differential equation, the pressure is obtained as
\begin{equation}
P=\frac{\alpha}{3}T^{4}+\gamma T\,,
\end{equation}
where 
\bea
\alpha &=& \frac{32\pi^4}{45(2M-r)^3(2Mr^3-r^4-4J^2\sin^2\theta)^2}\Bigl[8M^2r^9+2Mr^{10}-3r^{11}+88J^2Mr^6\sin^2\theta\nonumber\\
&-&44J^2r^7\sin^2\theta-72J^2Mr^9\sin^2\theta+36J^2r^{10}\sin^2\theta-64J^4r^3\sin^4\theta\Bigl]
\eea
and $\gamma$ is an integration constant. It is worth noting that equation (\ref{primeiralei}) is also called internal pressure. This procedure was the same used by Boltzmann to derive Stefan's law, for which the relationship between energy and pressure for electromagnetic waves was known. Indeed the chemical potential is assumed to vanish, but a gravitational statistical mechanics should be established in order to understand what are the respective micro-states. Since $\left(\frac{\partial P}{\partial T}\right)_{V} = \left(\frac{\partial S}{\partial V}\right)_{T}$, then
\begin{equation}
\left(\frac{\partial S}{\partial V}\right)_{T}=\gamma+\frac{4}{3}\alpha T^{3}\,,
\end{equation}
which leads to the following expression for the heat capacity at constant volume
\begin{equation}
c_{v}=4\alpha T^{2}\,,
\end{equation}
where $c_{v}=\frac{\partial^{2}S}{\partial T\partial V}$. It is observed that there is no phase transition for the temperature according to the Landau approach.  However, there is a spatial region in which the heat capacity is divergent, this occurs at $r=2M$ and at $2Mr^3-r^4-4J^2\sin^2\theta=0$.  It is therefore possible that there are distinct phases around the Kerr black hole linked to regions of spacetime. On the other hand, thermal capacity analysis is related to the $\alpha$ function. Thus we see the graphs of such a function when the parameters $M=J=0.1$, due to the limitations of our model, in figures FIG. 1, FIG. 2, FIG. 3 and FIG. 4. It is noticed that there is a divergence at only one point in the real plane, around $0.2$. It should be noted the negative behavior of $\alpha$ in a region when we fix the theta angle in the equatorial plane.

\begin{figure}[!thb] 
\caption{Behavior from $\alpha$ to $\theta=0$, with r from 0 to $0.2$.}
\includegraphics[scale=0.5]{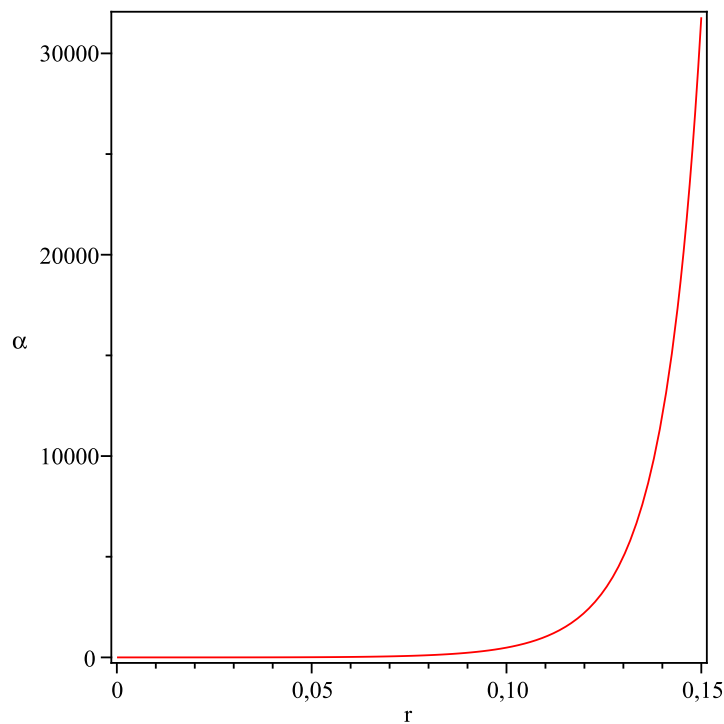} 
\end{figure}

\begin{figure}[!thb] 
\caption{Behavior from $\alpha$ to $\theta=0$, with r from 2 to $100$.}
\includegraphics[scale=0.5]{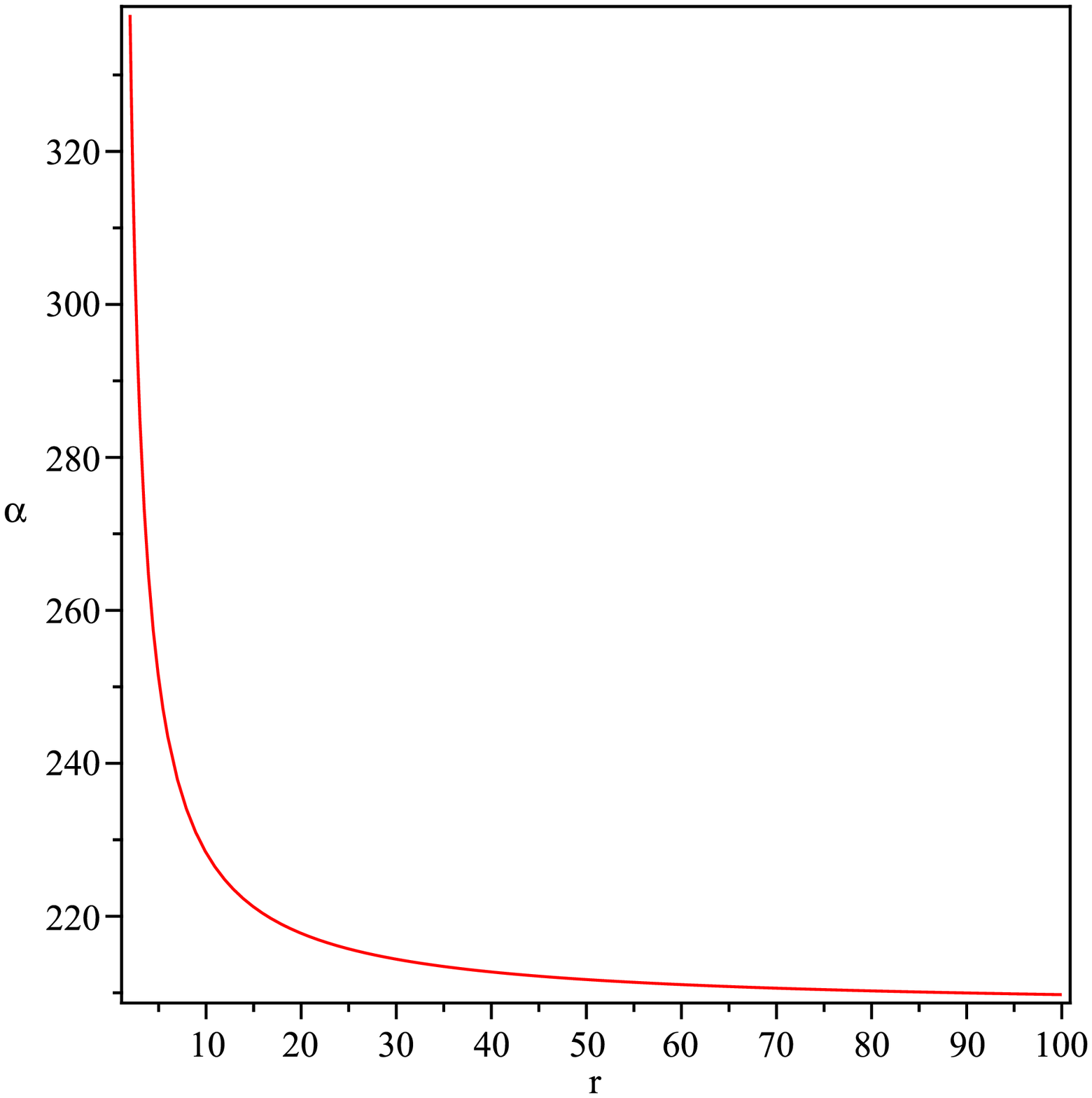} 
\end{figure}

\begin{figure}[!thb] 
\caption{Behavior from $\alpha$ to $\theta=\pi/2$, with r from 0 to $0.2$.}
\includegraphics[scale=0.5]{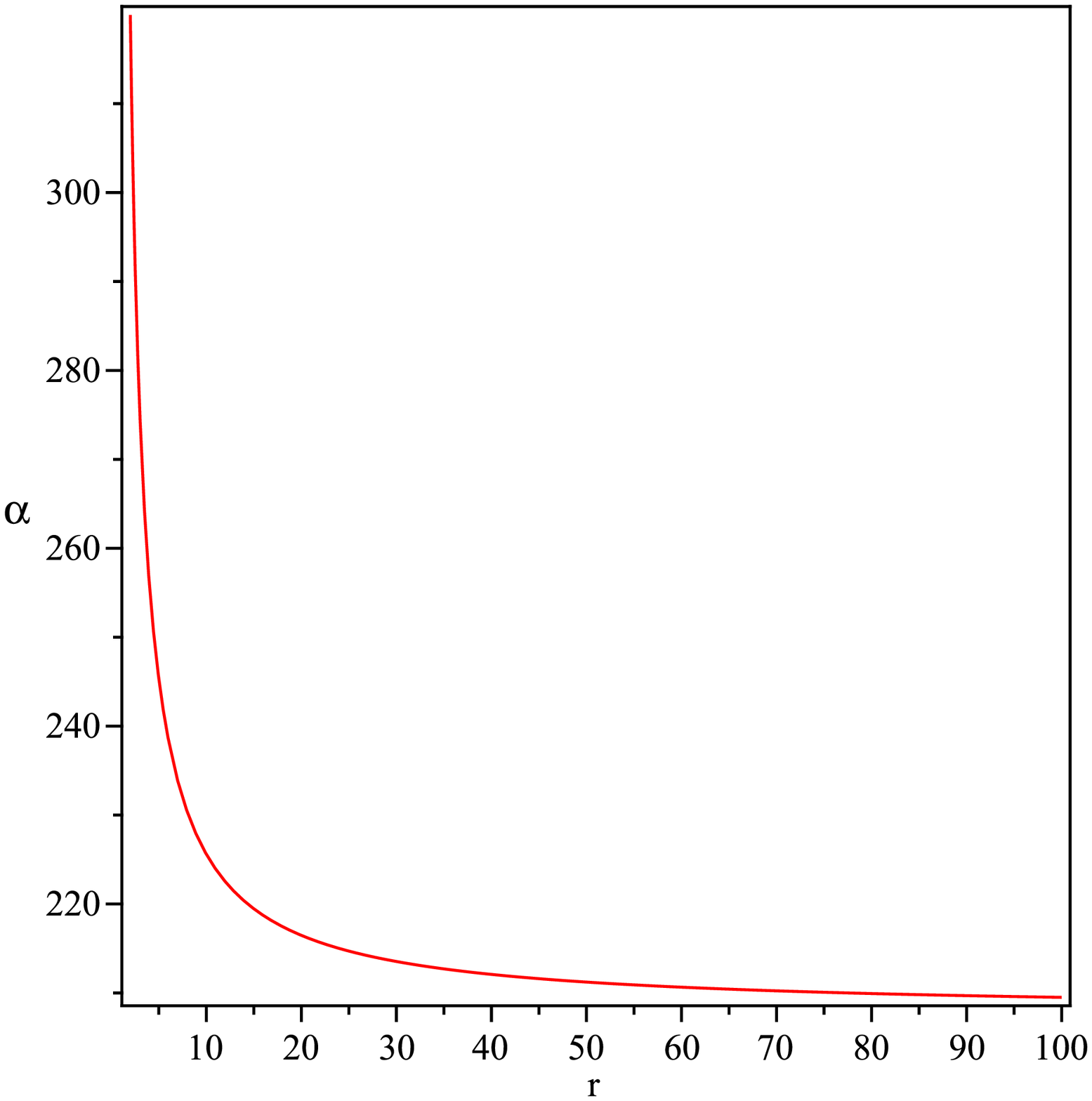} 
\end{figure}

\begin{figure}[!thb] 
\caption{Behavior from $\alpha$ to $\theta=\pi/2$, with r from 2 to $100$.}
\includegraphics[scale=0.5]{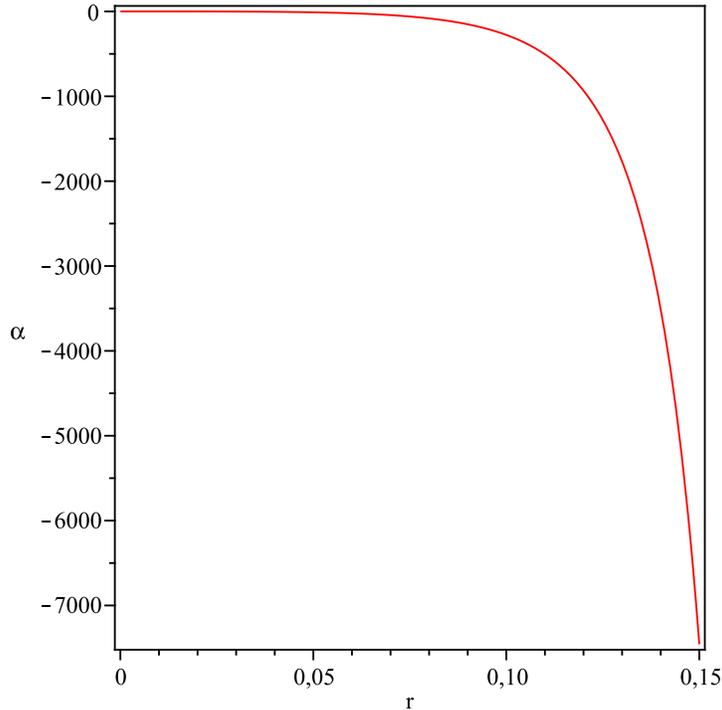} 
\end{figure}

\section{Conclusion} \label{sec.5}

In this paper, the gravitational Stefan-Boltzmann law is obtained for the Kerr black hole in a weak-field approximation.  From this, using the first law of thermodynamics, the gravitational pressure and the heat capacity are obtained.  These thermodynamic quantities are dependent on the physical parameters of the black hole, such as the mass and angular momentum, as well as temperature.  So it becomes an important part of Kerr black hole physics.  Unlike what usually happens for Kerr black hole thermodynamics, in which the temperature is dependent on the parameters of a black hole event horizon, here the temperature is entirely independent of the geometry.  However a possible spatial phase transition is observed. A spatial distribution of temperature can also occur in the Hawking approach, but it is obtained by an observer that is not stationary \cite{Brynjolfsson}. Therefore, a prediction of a temperature-dependent gravitational pressure with a very well-defined physical meaning is obtained.  Such a temperature must interact with a thermometer and thus lead to an experimental pattern, as usual thermodynamics establishes.  As far as it is known, this is a unique prediction of gravitational TFD. It is interesting to note that the $\gamma$ constant has a very clear interpretation.  For this, it must be realized that the pressure inflection temperature is, in fact, an equilibrium temperature.  This is because the entropy tends to a maximum value.  Since the $\alpha$ limit for $r$ tending to infinity is $\frac{32\pi^4}{15}$, then, at that limit, the following temperature is found
$$T_e =\left(-\frac{45\gamma}{128\pi^4}\right)^{1/3}\,.$$
This means that, at infinity, the Kerr black hole will exhibit a finite measurable temperature associated only to the gravitational field. Such a prediction contrasts with Hawking's approach that assigns a temperature only to the event horizon, although the thermal radiation could be measured at infinity. Truth be told, in this approach there is the possibility of having a temperature distribution beyond the event horizon if there is a coupling with an external field, but there is no case where a temperature is solely associated with gravitation outside the black hole \cite{Casals}. There is no reason to believe that this feature applies to other black holes as well.  So the experiment designed to measure gravitational waves can be altered to determine such a temperature, as the source of these waves is supposedly black holes.  It is interesting to note that in references \cite{prieto} and \cite{prieto2}, a new technique for estimating the temperature of black holes is presented.  In these references the authors discuss discuss an interesting effect, that is, the more massive the black hole, the colder it becomes.  This effect appears in the $\alpha$ function of the energy density predicted in this paper.  This is an indication that the $\gamma$ constant can be determined experimentally. On the other hand, there are no experimental results in the literature that indicate a temperature distribution of black holes, therefore, to confirm the prediction of $T_e$, a specific experiment must be designed.

\section*{Acknowledgments}

This work by A. F. S. is supported by National Council for Scientific and Technological Develo\-pment - CNPq project 313400/2020-2.


\begin{thebibliography}{99}

\bibitem{dinverno}

R. D'Inverno. Introducing Einstein's Relativity. Clarendon Press, 1992.

\bibitem{waves}

B. P. Abbott et al. (LIGO Scientific Collaboration and Virgo Collaboration)
Phys. Rev. Lett. {\bf 116}, 061102 (2016).

\bibitem{photo}

M. Wielgus et al. Astrophys. J. {\bf 901}, 67 (2020).

\bibitem{hawking}

S.W. Hawking, Physical Review Letters, {\bf 26}, 1344 (1971).

\bibitem{Bekenstein} 

J.~D. Bekenstein. Black Holes and Entropy. Physical Review D, 7(8):2333, (1973).

\bibitem{Jacobson}

T. Jacobson. Thermodynamics of Spacetime: The Einstein Equation of State. Physical Review Letters, {\bf 75}, 1260, (1995).

\bibitem{Padmanabhan}

T. Padmanabhan. Thermodynamical Aspects of Gravity: New insights. Rep. Prog. Phys., {\bf 73}, 046901, (2010).

\bibitem{Isi}

M.~Isi, W.~M Farr, M.~Giesler, M.~A. Scheel, and S.~A. Teukolsky. Testing the black-hole area law with gw150914. Physical Review Letters, {\bf 127}, 011103, (2021).

\bibitem{Wald}

R.~M. Wald. The thermodynamics of black holes. Living reviews in relativity, {\bf 4}, 1-44, (2001).

\bibitem{dolan}

B.P. Dolan, ``Where is the PdV in the first law of black hole thermodynamics?'' in Open Questions in Cosmology, chapter 12, (2012).

\bibitem{matsubara}

T. Matsubara, Prog. Theor. Phys. {\bf 14}, 351 (1955).

\bibitem{khanna}

Faqir C Khanna, Adolfo P C Malbouisson, Jorge M C Malbouisson and Ademir E Santana. Thermal Quantum Field Theory. WORLD SCIENTIFIC, 2009.


\bibitem{maluf}

J. W. Maluf. The teleparallel equivalent of general relativity. Annalen der Physik, {\bf 525}(5):339–357, 2013.


\bibitem{flatTFD}

S. C. Ulhoa, A. F. Santos and F. C. Khanna, General Relativity and Gravitation {\bf 49},  54, (2017).

\bibitem{Umezawa1}Y. Takahashi and H. Umezawa, Collective Phenomena 2, 55 (1975) (Reprinted in Int. J. Mod. Phys. B 10, 1755 (1996)).
\bibitem{Umezawa2}Y. Takahashi, H. Umezawa and H. Matsumoto, Thermofield Dynamics and Condensed States, North-Holland, Amsterdan, (1982).
\bibitem{Khanna0} F. C. Khanna, A. P. C. Malbouisson, J. M. C. Malboiusson and A. E. Santana, Themal quantum field theory: Algebraic aspects and applications, World Scientific, Singapore, (2009).
\bibitem{Umezawa22} H. Umezawa, Advanced Field Theory: Micro, Macro and Thermal Physics, AIP, New York, (1993).
\bibitem{Khanna1} A. E. Santana and F. C. Khanna, Phys. Lett. A {\bf 203}, 68 (1995).
\bibitem{Khanna2} A. E. Santana, F. C. Khanna, H. Chu, and C. Chang, Ann. Phys. {\bf 249}, 481 (1996).
\bibitem{GBT}F. C. Khanna, A. P. C Malbouisson, J. M. C. Malbouisson and A. E. Santana, Ann. Phys. {\bf 326}, 2634 (2011).

\bibitem{ulhoa}

S. C. Ulhoa, F. L. Carneiro, and J. W. Maluf, Modern Physics Letters A {\bf 37}, 32, 2250219 (2022).

\bibitem{Brynjolfsson}

Erling J. Brynjolfsson and Larus Thorlacius, JHEP09 {\bf 2008}, 066 (2008).



\bibitem{Casals}

M. Casals, S. R. Dolan, B. C. Nolan, A. C. Ottewill and E. Winstanley,
Phys. Rev. D {\bf 87}, 064027 (2013).

\bibitem{prieto}

A. Prieto,  A. Rodr\'iguez-Ardila,  S. Panda and M. Marinello, Mon. Not. Royal Astro. Soc., {\bf 510}, 1, 1010 (2022).

\bibitem{prieto2}

A. Prieto,  A. Rodr\'iguez-Ardila,  S. Panda and M. Marinello, Mon. Not. Royal Astro. Soc., {\bf 511}, 3, 4444 (2022).



\end{thebibliography}
\end{document}